\documentclass{article}
\usepackage[a4paper, margin=1in]{geometry}
\usepackage{cite}
\usepackage{graphicx} 
\usepackage{amsmath}
\usepackage{amssymb}
\usepackage[version=4]{mhchem}
\usepackage{float}
\usepackage{authblk}

\title{Measurement of complex scattering matrix in a nano-cavity array for boundary scattering tomography}

\author[1,*]{Andrew Tang}
\author[1]{Romil Audhkhasi}
\author[1]{Virat Tara}
\author[1]{Abhi Saxena}
\author[1]{Gokul Nath}
\author[1,2]{Arka Majumdar}

\affil[1]{Department of Electrical and Computer Engineering, University of Washington, Seattle, Washington 98195, United States}

\affil[2]{Department of Physics, University of Washington, Seattle, Washington 98195, United States}

\affil[*]{adtang@uw.edu}

\date{}

\begin{document}

\maketitle

\section{Abstract}

On-chip silicon photonic coupled cavity arrays (CCA) are a promising platform for quantum simulators, with access to high Quality (Q) factor resonators, tunability, and foundry compatibility. Furthermore, scalable two-dimensional (2D) silicon photonic CCAs allow for simulation of rich physical phenomena via Hamiltonian engineering. However, complete reconstruction of the Hamiltonian is limited by access to cavities in the bulk, with current approaches relying on imaging scattered light from bulk resonators. These approaches often require additional scatterers to be built in, limiting scalability, while also being hampered by imaging technology in the near-infrared range. Instead of these approaches, Hamiltonian tomography algorithms that require homodyne boundary measurements have been demonstrated in literature, however measurements of complex scattering measurements along a CCA boundary have not been shown. Here, we experimentally demonstrate an on-chip homodyne measurement setup along a single boundary of a $3\times 3$ silicon photonic racetrack resonator array and reconstruct the system's edge scattering matrix.

\section{Introduction}

Photonic coupled-cavity arrays (CCAs) have emerged as a promising platform for analog quantum simulation due to their intrinsic scalability, tunability, and long coherence time of photons~\cite{georgescu_quantum_2014, aspuru-guzik_photonic_2012, hartmann_quantum_2008}. In particular, silicon photonics enable tightly confined high-$Q$ optical resonators with independently tunable elements, offering precise control over on-site energies and inter-cavity couplings, all while being compatible with existing high volume semiconductor manufacturing processes~\cite{adcock_advances_2021}. One dimensional (1D) CCAs have been extensively explored in coupled-resonator optical
waveguides (CROW), and recent demonstrations in 1D CCAs have implemented programmable tight-binding Hamiltonians on-chip, with up to eight coupled silicon microrings whose full eigenmode spectra can be accessed and tuned in situ~\cite{poon_matrix_2004, tang_cavitywaveguide_2025, saxena_realizing_2023}.

These demonstrations showcase the potential of this platform, and scaling from one- to two-dimensional arrays is the natural next step. To this end, 2D CCAs have been actively explored for topological phemonena, including seminal works demonstrating fractional quantum hall effects and robust edge states ~\cite{hafezi_imaging_2013, li_topological_2022, gao_topological_2023, dai_programmable_2024}. Furthermore, different platforms for 2D CCAs have demonstrated the simulation of complex models such as the 2D Su–Schrieffer–Heeger model and graphene~\cite{kremer_demonstration_2019, zheng_observation_2019}. However, in all these structures, complete Hamiltonian tomography, i.e., estimating all parameters of the Hamiltonian, was not reported. As such, extending on-chip photonic lattice simulators from 1D to 2D poses new challenges in terms of measurement and scalability. In a 2D CCA, resonators in the bulk cannot be directly accessed by input/output ports, making it difficult to probe or address individual sites beyond the edges. Conventional approaches monitor every cavity by imaging scattered light from each resonator, often through the inclusion of scatterers, which incur substantial hardware overhead while increasing optical losses. Furthermore, for silicon cavities operating at telecommunication wavelengths ($1310nm$ or $1550nm$), imaging technology for collecting scattered light lags behind, especially compared to those found in the visible or near-visible~\cite{sharp_near-visible_2024}. As a result, prior experiments in on-chip 2D photonic lattices have either been limited to observing only edge phenomena or requiring elaborate imaging techniques to infer internal light distributions.

\begin{figure}[H]
    \centering
    \includegraphics[width=0.8\columnwidth]{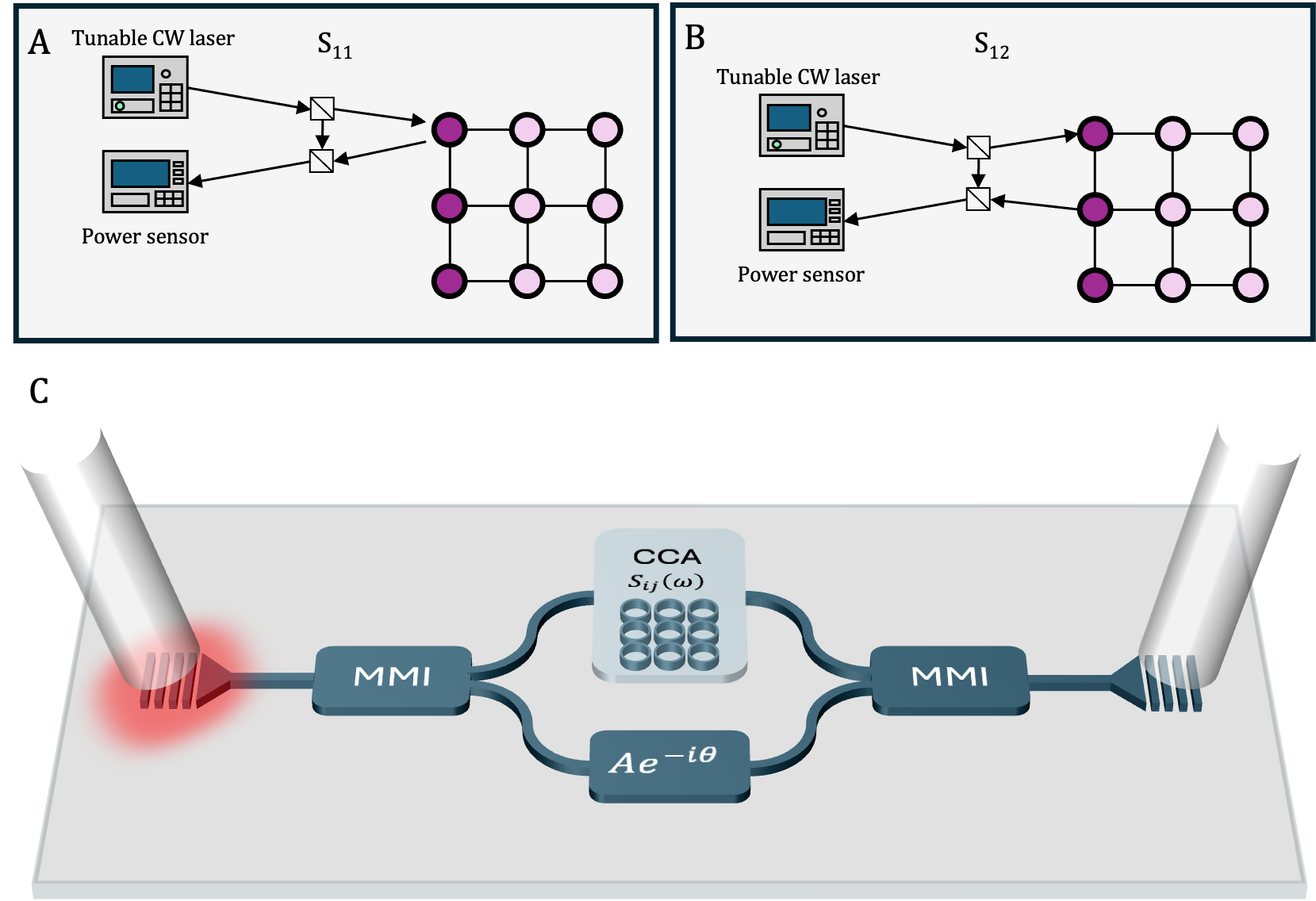}
    \caption{Complex scattering matrix measurement along a single edge of a 2D CCA array. Dark purple circles represent cavities along the measurement edge, while light circles represent cavities in the bulk. Dotted lines represent cavity cavity to cavity coupling, in this case nearest neighbors are coupled. A) Homodyne measurements with same input and output cavity, i.e. $S_{11}$. B) Homodyne measurements with input and output cavities being adjacent, i.e. $S_{12}$. C) Schematic of on-chip interferometric measurement setup, with reference arm, $Ae^{-i\theta}$, and CCA arm, $S_{ij}(\omega)$.}
    \label{fig:design}
\end{figure}

To overcome these limitations, we employ boundary-only, phase-resolved measurements of the complex scattering matrix using on-chip homodyne interferometry. Rather than probing cavities in the bulk, recent Hamiltonian tomography algorithms have shown reconstruction of the tight-binding Hamiltonian (all on-site frequencies and coupling rates) through boundary complex scattering matrices~\cite{saxena_boundary_2024}. This algorithm requires measurements across a single cavity along the boundary (e.g., $S_{11}$) as well as between adjacent cavities along the boundary (e.g., $S_{12}$), as illustrated in Fig.~\ref{fig:design}A–B. Crucially, implementing these measurements on chip avoids phase noise and unwanted interference fringing arising from fiber jitter and path-length mismatch in off-chip interferometric setups. In this work, we demonstrate an integrated homodyne measurement architecture for a $3 \times 3$ silicon photonic CCA and reconstruct the complex scattering matrix across a single edge boundary.

\section{Theory}

In this work, we consider a 2D CCA in a square lattice comprised of identical racetrack resonators with equal designed coupling rates between resonators. The effective Hamiltonian of the CCA is

\begin{equation}
    H_{eff} = \sum_{i=1}^{N} \sum_{j=1}^{N} (\mu-i\kappa_c) a_{i,j}^{\dagger}a_{i,j} - J \sum_{i=1}^{N-1} \sum_{j=1}^{N} (a_{i + 1,j}^{\dagger}a_{i,j}+\text{h.c.}) - J\sum_{i=1}^{N} \sum_{j=1}^{N-1} ( a_{i ,j + 1}^{\dagger}a_{i,j} + \text{h.c.}) - i\gamma_{c} \sum_{i=1}^{N}  a_{i,1}^{\dagger}a_{i,1}
\end{equation}

where $\mu$ is the detuning between the cavity resonance frequency and the driving laser, $J$ is the complex nearest neighbors field coupling rate between resonators, $\gamma_{c}$ is the waveguide to cavity field coupling rate along the bottom row, and $\kappa_c$ denotes the intrinsic field loss of the resonators. In the identical resonator and coupling regime, $J$ is fully real, however our result generalizes to complex $J$ values.

This CCA can further be modeled through a complex scattering matrix, whereas the complex transmission with input resonator $i$ and output resonator $j$ can be modeled as a sum of Lorentzians,

\begin{equation}
    S_{ij}(\omega) = \delta_{ij} - i\times\gamma_{c}\sum_{k=1}^{N^2} \frac{M_{ij, k}}{\omega-E_{k}}
\end{equation}

in which $\delta_{ij}$ is the Kronecker delta, $E_k$ represents the complex energies of $H_{eff}$, and $M_{ij,k}$ captures the overlap between the eigenstates of $H_{eff}$ and boundary-localized excitations~\cite{saxena_boundary_2024}. To recover $S_{ij}$, we consider an interferometer setup as shown in Fig.~\ref{fig:design}C with measurable function

\begin{equation}
    T_{ij}(\omega, \theta) = \left|A_{\mathrm{referece}}\times e^{-i\theta} +A_{\mathrm{CCA}}\times S_{ij}(\omega)\right|^2
\end{equation}

where $A_{\mathrm{referece}}$ and $A_{\mathrm{CCA}}$ are the E-field amplitudes in the reference and CCA arm of the interferometer, respectively, $\omega$ is the angular frequency of the light, and $\theta$ represents the phase difference between arms.

\section{Results}

The proposed system consists of a $3\times 3$ array of racetrack resonators, shown in Fig.~\ref{fig:setup}A and Fig.~\ref{fig:setup}C, and multiple MZIs for homodyne measurements, shown in Fig.~\ref{fig:setup}A. The array of resonators is composed of nine equal-length racetrack resonators, with straight side lengths of 8 $\mu$m and bending radii of 5 $\mu$m. Resonators are evanescently coupled along their straight edges with a 70 nm gap resulting in a theoretical coupling rate of approximately $131.0 \times 2\pi$ GHz (see supplemental materials). Each resonator is uniformly spaced along a square lattice such that there is no offset between adjacent resonators. These design choices lead to a degenerate effective Hamiltonian with five modes rather than nine.

\begin{figure}[H]
    \centering
    \includegraphics[width=0.75\columnwidth]{fig2.png}
    \caption{A) The complete architecture consisting of 2D CCA and the phase measurement structures, with input and output ports labeled as well as MZI reference arms labeled based off S-measurement. The critical parts of the architecture highlighted in: B) Photonic attenuator based off tunable MZI to allow for non-interference measurement. C) Optical micrograph of the $3\times 3$ silicon photonic coupled cavity array (CCA) made of racetrack resonators. D) 1 by 3 multi-mode interference coupler (MMI) for splitting light to an edge cavity, the output of that edge cavity, and the output of the adjacent cavity.  E) Thermo-optic phase tuner using platinum (Pt) heater for controlling phase difference between MZI arms.}
    \label{fig:setup}
\end{figure}

The phase-coherent measurement architecture is based on a balanced Mach-Zehnder Interferometer (MZI) as shown in Fig.~\ref{fig:design}C, with a multi-mode interference coupler (MMI) acting in the role of the canonical two way mirror to split the light between two arms. One arm leads to the CCA with a fixed complex scattering matrix $S_{ij}(\omega)$, whereas the other arm serves as a reference arm with a tunable response $A_{reference}e^{-i\theta}$. Through varying $\theta$ and observing the measurable $T_{ij}(\omega,\theta)$, we are able to use Equation 3 to reconstruct the complex scattering matrix $S$ by measuring across a boundary. 

To achieve this, we need measurement across the same cavity, i.e. Fig.~\ref{fig:design}A, as well as different cavities, i.e. Fig.~\ref{fig:design}B, therefore we use a 1 by 3 MMI as shown in Fig.~\ref{fig:setup}D. One MMI output is the CCA arm with designed coupling strength of $15.92 \times 2\pi$ GHz, which is much less than the cavity-cavity coupling strength of the CCA [see supplemental materials] to keep the CCA dynamics unaffected. The other two MMI outputs serve as reference arms, with one leading to the output port of the input cavity and the other leading to the output port of the adjacent cavity. In addition, the path lengths of all three arms are matched using switchbacks, as shown in Fig.~\ref{fig:setup}A, to avoid interference fringing. The phase difference, $\theta$ between the reference and CCA arm is controlled using a thermo-optic heater, as seen in Fig.~\ref{fig:setup}E, while the amplitude of the reference signal, $A_{reference}$ is controlled using an MZI based attenuators as shown in Fig.~\ref{fig:setup}B. The attenuator is necessary for cross resonator measurements, i.e. $S_{12}$, to ensure both arms in the MZI have comparable signal strength. The combination of these elements allows for the phase-coherent, boundary measurements necessary for complex scattering matrix measurements. The complex scattering matrix would then allow for Hamiltonian tomography, which however  is precluded due to the mode degeneracy.

\begin{figure}[H]
    \centering
    \includegraphics[width=0.7\columnwidth]{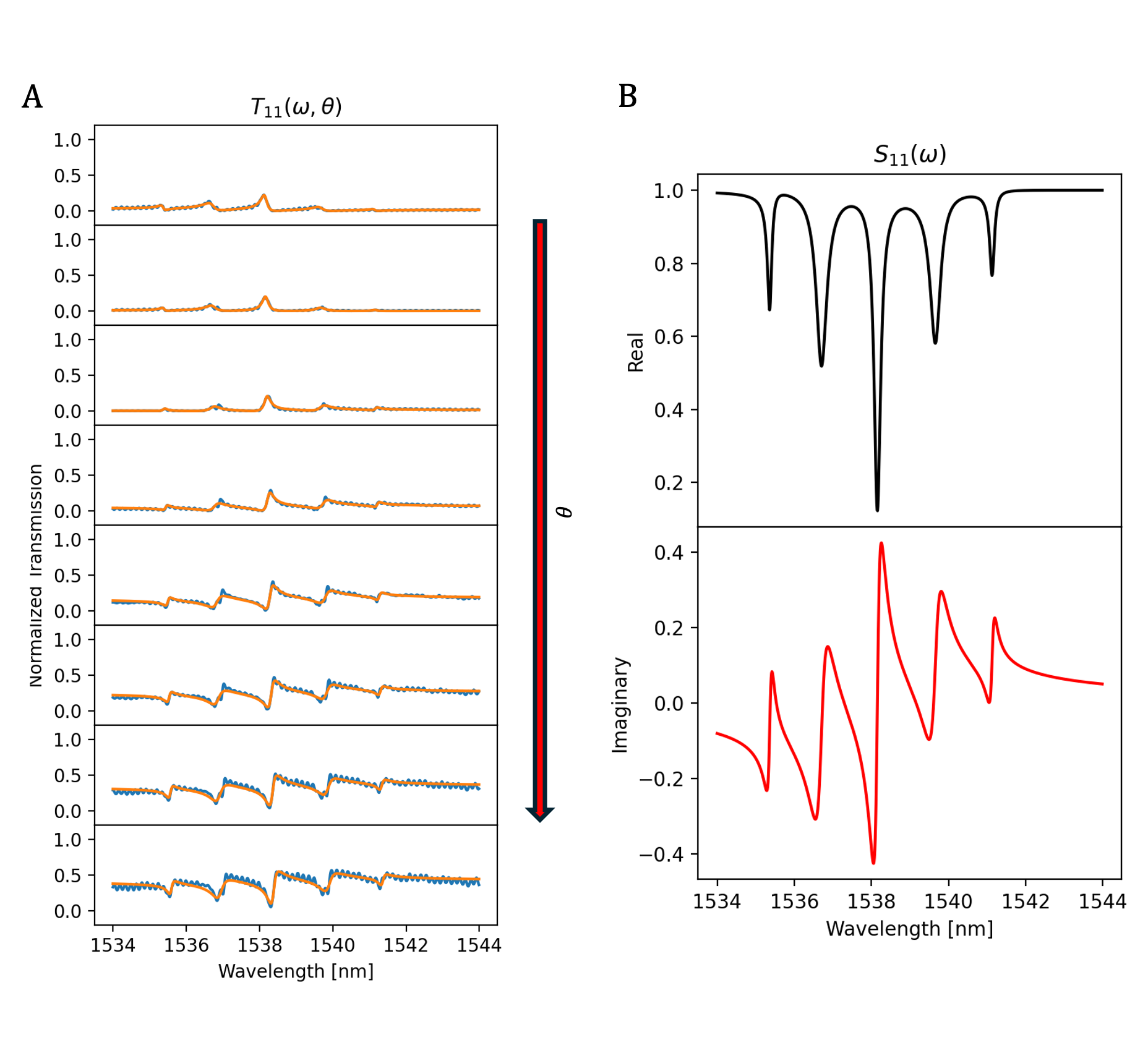}
    \caption{A) $S_{11}$-port measurements on the outer edge of the CCA  in blue with fitting in orange as a function of phase difference between MZI arms, $\theta$. B) Normalized fitted complex transmission with real and imaginary components plotted separately.}
    \label{fig:experimental}
\end{figure}

\begin{table}[htbp]
\centering
\caption{Fitted resonance parameters for S11.}
\begin{tabular}{llll}
\hline
$M_{\mathrm{real}}\,[10^{-2}]$ & $M_{\mathrm{imag}}\,[10^{-3}]$ & $E_{\mathrm{real}}$ & $E_{\mathrm{imag}}$ \\
\hline
$2.15 \pm 0.06$   & $-4.7 \pm 0.7$ & $1535.483 \pm 0.003$ & $0.0681 \pm 0.0025$ \\
$8.00 \pm 0.11$   & $2.3 \pm 1.2$  & $1536.819 \pm 0.003$ & $0.1674 \pm 0.0025$ \\
$8.90 \pm 0.07$   & $2.7 \pm 0.8$  & $1538.269 \pm 0.001$ & $0.1031 \pm 0.0007$ \\
$6.77 \pm 0.12$   & $-6.4 \pm 1.2$ & $1539.782 \pm 0.003$ & $0.1654 \pm 0.0031$ \\
$1.71 \pm 0.08$   & $-2.3 \pm 0.8$ & $1541.251 \pm 0.004$ & $0.0747 \pm 0.0042$ \\
\hline
\end{tabular}
\label{tab:resonance_params}
\end{table}

\begin{figure}[H]
    \centering
    \includegraphics[width=1.0\columnwidth]{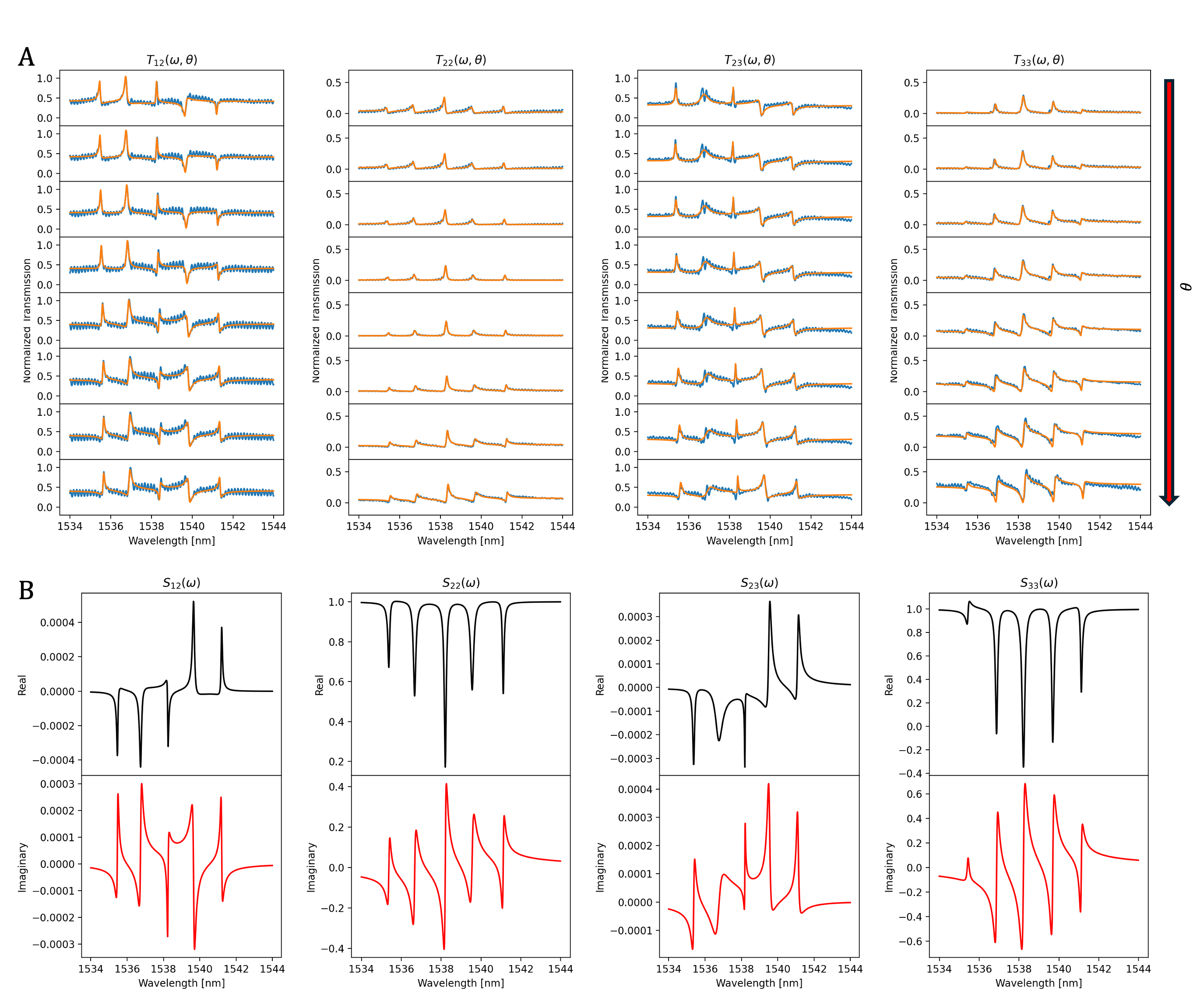}
    \caption{A) Remaining S-port measurements along a single edge of the CCA as a function of phase difference between MZI arms. B) Normalized fitted complex transmission with real and imaginary components plotted separately.}
    \label{fig:alldata}
\end{figure}

To achieve the complex boundary scattering matrix along a single edge, we take five sets of spectrum measurements: $T_{11}(\omega, \theta)$, $T_{12}(\omega, \theta)$, $T_{22}(\omega, \theta)$, $T_{23}(\omega, \theta)$, and $T_{33}(\omega, \theta)$ with $T_{11}$ shown in Fig.~\ref{fig:experimental}A and the others in Fig.~\ref{fig:alldata}A. All measurements are taken within the free spectral range (FSR) of a single racetrack resonator which we experimentally find to be 1030 GHz [8.2 nm], such that we function with a single mode approximation with no overlap between higher and lower order modes. Each port configuration has fixed complex transmission parameters $E$ and $M$ shown in Equation 2, which are globally fit for all transmission measurements of that port, such as those shown in Fig.~\ref{fig:experimental}A, with the only variation within being the phase difference between MZI arms, $\theta$ (see supplementary materials). The phase difference is changed through tuning the voltage across the thermo-optic heater on the CCA arm of the MZI from 0 to 3.4V. The fitted complex transmission for $S_{11}$ is shown in Fig.~\ref{fig:experimental}B with the real and imaginary parts separated. Each peak in Fig.~\ref{fig:experimental}B corresponds to a complex Lorentzian with the fitted parameters of the corresponding peak and fitting error shown in Table 1 for the $M_{11, i}$ and $E_i$ components respectively. Whereas the fitted $M$ parameters are local to the specific measured resonator, the $E$ parameters correspond to the complex energies of $H_{eff}$ for the whole system.

In the demonstrated system with identical resonators and identical couplings, the complex energies of $H_{eff}$ can be found as the eigenvalues of the matrix representation of $H_{eff}$ shown in Equation 1. Taking the eigenvalues of the theoretical system, there are five complex energies at $\mu-i\kappa_c$, $(\mu-i\kappa_c)\pm\sqrt{2}J$, and $(\mu-i\kappa_c)\pm2\sqrt{2}J$ [see supplemental materials], corresponding to the five resonances shown in Fig.~\ref{fig:experimental}B. Taking the real distance between adjacent resonances obtained through the fitting, we can estimate the experimental $J = 128.9\pm6$ GHz, which closely matches our designed theoretical value of $131.0$ GHz. Furthermore, because the imaginary component of $E$ corresponds to the linewidth of the supermodes of the CCA, we estimate a lower bound on the intrinsic Q of individual racetrack resonators. Taking the minimal supermode linewidth from Fig.~\ref{fig:experimental}D, yields an intrinsic Q factor $\gtrsim1.3\times10^4$. This approximation is for the lowest loss resonator, however we expect comparable intrinsic losses across resonators and furthermore the inferred value is likely an underestimation due to the added loss from boundary resonators due to waveguide coupling.

Fig.~\ref{fig:alldata}A shows the rest of the S-parameters along a single edge of the 3 x 3 2D CCA, with measurements between adjacent cavities, as a function of phase difference between MZI arms. The fitted complex transmission is shown in Fig.~\ref{fig:alldata}B.

\section{Conclusion}
In this work, we have demonstrated measurement of the complex transmission along the edge of a 2D coupled cavity array of 9 racetrack resonators through the usage of phase-coherent, boundary measurements. From these complex measurements, we are able to recover $E$, the complex energies  of the effective Hamiltonian, and $M$, the eigenstate overlaps of the measured resonator with $H_{eff}$. Furthermore, our measurements and fittings show consistency throughout all the different port measurements and closely match the theoretical coupling rates. These lattice measurements serve as a proof of concept for an on chip measurement setup necessary for Hamiltonian tomography algorithms that could serve as a path towards creating large scale, photonic quantum simulations.

\section{Methods}
\subsection{Device Fabrication}
The device was patterned using electron beam lithography (JEOL JBX6300FS) on a silicon-on insulator (SOI) platform consisting of 220 nm of Si on 2 $\mu$m of SiO2. The resist used was hydrogen silsesquioxane (HSQ) resist and development was done using TMAH 25\%. 220 nm of Si was etched with Cl2 gas and a BOE wet etch was used afterwards to remove leftover resist. Then, 1 $\mu$m of SiO2 was grown using PECVD to allow for metal heaters, which are patterned using the Heidelberg DWL66+ photolithography tool with AZ1512 resist. Then, electron beam evaporation was used to deposit 25 nm of Ti for adhesion and 325 nm of Pt for heating. Lastly, heaters are wirebonded to a chip carrier for DAQ control.

\subsection{Device Characterization}
The device spectrum was characterized via a fiber array setup using a tunable continuous-wave laser (Santec TSL-510) excitation to the input grating and a low-noise power meter (Keysight 81634B) measurement of the output grating. Thermo-optic tuning of the resonators and waveguides was achieved using a DAQ (MCC USB 3114).

\subsection{Device Simulations}
Commercial softwares Ansys Lumerical FDTD, MODE, HEAT, and INTERCONNECT were used to simulate and optimize the device parameters. Lumerical FDTD and MODE were used to find the necessary resonator and waveguide dimensions to achieve the desired coupled mode parameters.

\section{Acknowledgments}
We would like to thank Rahul Trivedi for the useful discussions on Hamiltonian tomography algorithms. This paper is based predominantly on work conducted at the University of Washington supported by National Science Foundation (NSF) Science and Technology Center (STC) for Integration of Modern Optoelectronic Materials on Demand (IMOD) under Cooperative Agreement No. DMR-2019444. Part of work on sample fabrication and characterization was conducted at the Washington Nanofabrication Facility / Molecular Analysis Facility, a National Nanotechnology Coordinated Infrastructure (NNCI) site at University of Washington with partial support from NSF via awards NNCI-1542101 and NNCI-2025489.

\newpage
\bibliography{references}
\bibliographystyle{ieeetr}
\end{document}